\documentstyle[aps,manuscript,epsf]{revtex}
\begin{document}

\title{Direct search for solar axions by using strong magnetic
field and X-ray detectors}

\author{\underline{Shigetaka~Moriyama}$^a$
\thanks{E-mail: moriyama@icepp.s.u-tokyo.ac.jp},
Makoto~Minowa$^a$,
Toshio~Namba$^a$,\\
Yoshizumi~Inoue$^b$,
and Yuko~Takasu$^a$}
\address{
$^a$Department of Physics, School of Science,
University of Tokyo,\\
7-3-1 Hongo, Bunkyo-ku, Tokyo 113-0033, Japan\\
$^b$International Center for Elementary Particle Physics,
University of Tokyo,\\
7-3-1 Hongo, Bunkyo-ku, Tokyo 113-0033, Japan}
\author{Akira~Yamamoto$^c$}
\address{$^c$High Energy Accelerator Research Organization (KEK),\\
1-1 Oho, Tsukuba, Ibaraki 305-0801, Japan}

\maketitle

\begin{abstract}
We have searched for axions which could be produced
in the solar core by exploiting their conversion to X rays
in a strong laboratory magnetic field.
The signature of the solar axion is an increase
in the rate of the X rays detected in a magnetic helioscope
when the sun is within its acceptance.

From the absence of such a signal we set a 95\% confidence
level limit on the axion coupling to two photons
$g_{a\gamma\gamma}\equiv 1/M < 6.0\times 10^{-10}$\,GeV$^{-1}$,
provided the axion mass $m_a<0.03$\,eV.
The limit on the coupling is factor 4.5 more stringent
than the recent experimental result.
This is the first experiment whose sensitivity
to $g_{a\gamma\gamma}$ is higher than
the limit constrained by the solar age consideration.

\end{abstract}
\newpage

\section{Introduction}
In the standard model of the elementary particle physics,
it is thought that the strong interaction is
well described by the quantum chromodynamics.
However, the strong $CP$ problem which results from
the solution of $U(1)_A$ problem is still an open question.
Although Peccei and Quinn gave an attractive solution
\cite{PQ77} to the strong $CP$ problem,
the predicted particle \cite{WW},
axion, is not discovered yet.
It was originally thought that the Peccei-Quinn (PQ)
symmetry breaking, which was introduced to
explain the absence of $CP$ violating effects
in strong interactions,
occurs at the weak interaction energy scale,
$f_{PQ} \sim 250$\,GeV. The existence of axions
in this region has been experimentally excluded
\cite{Physrep}.

Axions that couple directly to electrons \cite{DFSZ}
through an $eea$ vertex provide a very efficient
energy-loss mechanism and the coupling
is strictly constrained by the cooling
rates of the sun, the red giants, and the supernova 1987A.
However, ``hadronic axions,'' which do not directly couple
to leptons \cite{KSVZ},
interact with matter primarily through a two-photon vertex.
They can still be produced in the solar core.
Such axions, if they exist, must be produced abundantly
in the core of the sun through
the Primakoff effect of black body photons.
They are referred to as solar axions.
They could be detected by their reconversion
into X rays in a strong laboratory magnetic field
\cite{Sikivie83,Bibber89}.
The expected energy of the X rays is distributed over 2--20\,keV.

The axion luminosity should be less than the corresponding photon
luminosity so as not to conflict with the apparent age of the sun;
otherwise its nuclear fuel would have been spent out
before reaching an age of $4.5\times 10^9$\,yr.
This requirement yields a bound to
the coupling constant to photons
$g_{a\gamma\gamma} \equiv 1/M < 2.4\times 10^{-9}\,\rm GeV^{-1}$.
Although there were several experiments
so far in which solar axions are searched for,
none have obtained sufficient sensitivity
to detect such weakly interacting axions.

A recent astrophysical argument on the horizontal-branch
stars in the globular clusters
by Raffelt \cite{RaffeltBook} further constrain
$g_{a\gamma\gamma}$ to be less than $6\times 10^{-11}$\,GeV$^{-1}$.
However, a direct observational search for solar axions
with sensitivity $g_{a\gamma\gamma} < 2.4\times 10^{-9}$\,GeV$^{-1}$
should have a qualitatively different meaning on the axion physics.
Possible loophole in the hypotheses
would be much less probable in the direct
observational experiment.

In this paper, we will describe
an experiment to search for solar axions.
For the first time, we obtained better sensitivity
than that obtained by the solar age consideration.

\section{Experimental Apparatus}
\label{sec:experimentalapparatus}
\begin{figure}
  \begin{center}
    \leavevmode
    \epsfxsize=15cm
    \epsffile{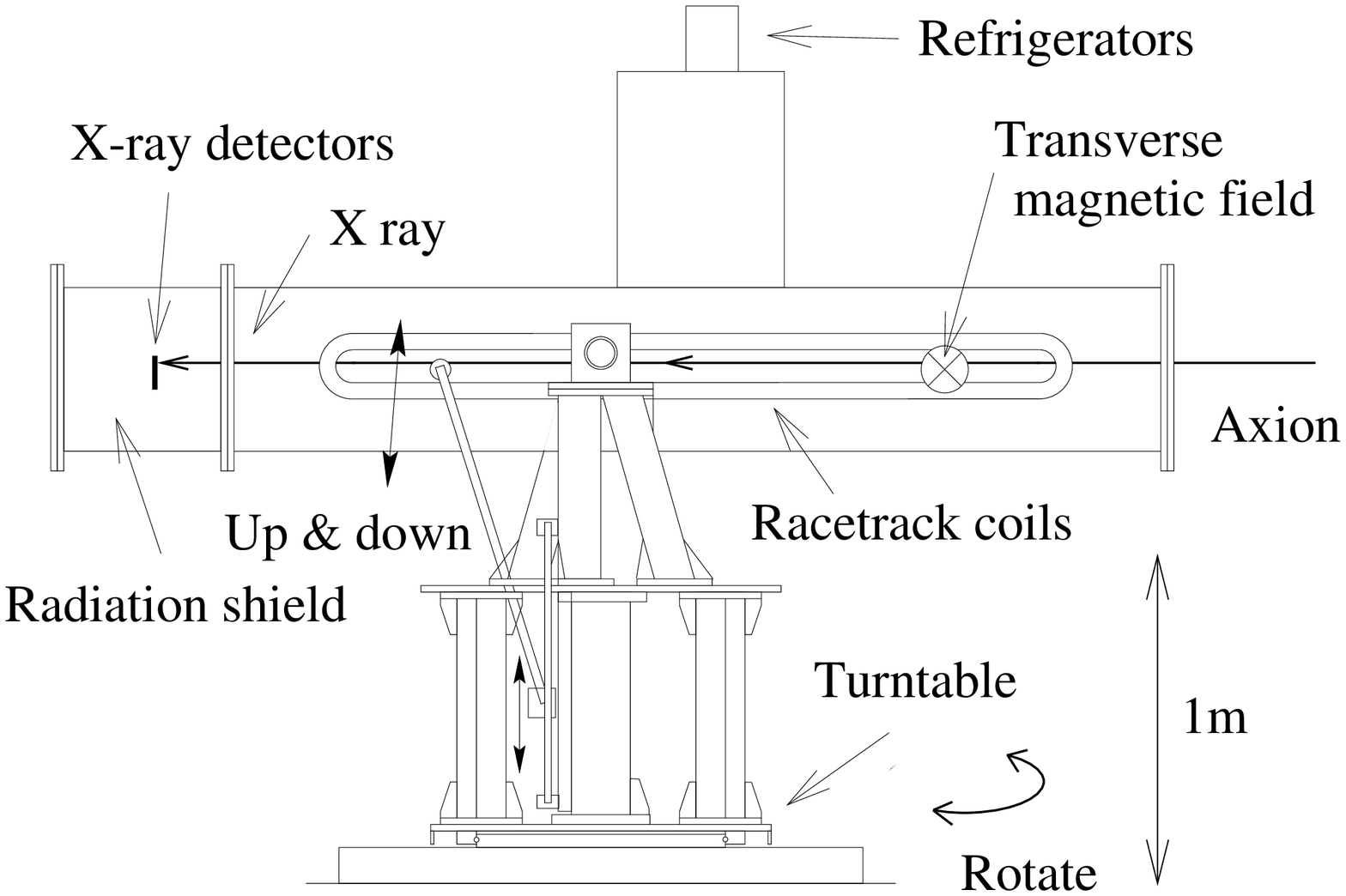}
    \caption{Schematic view of the apparatus}
    \label{fig:Sumiko}
  \end{center}
\end{figure}

We have constructed a superconducting magnet to obtain
a strong magnetic field, a turntable to direct the magnet
to the sun, and X-ray detectors. 
In Fig.\ \ref{fig:Sumiko}, schematic view of the apparatus is shown.
The cylindrical vacuum chamber contains
the superconducting magnet, X-ray detectors,
and a radiation shield to reduce background events.
In the bottom of the figure, the turntable is indicated.
It is designed to move the detector toward the sun
with sufficient accuracy.

The superconducting magnet consists of two split racetrack coils
whose length is 2.3\,m, width is 50\,mm,
and thickness is 64\,mm \cite{Sato}.
A dipole magnetic field generated by the two coils
is perpendicular to the pass of the axions
as shown in Fig.\ \ref{fig:Sumiko}.
Between two coils, there is a vacuum bore
in which the axions are expected to be converted into photons.
The width and height of the bore are 20\,mm and 92\,mm, respectively.
The magnetic field strength inside the bore
was kept to be stronger than 3.93\,T.
The coils are directly cooled down by two
Gifford-McMahon refrigerators
with a thermal conductive rod
and kept at 5.2\,K during the measurement.
A persistent current switch and current leads
are also placed on the top of the rod.
There are two copper plates which
sandwich the coils from above and below
to obtain a good thermal conductivity along the long axis.

Since we must move the magnet to track the sun,
the following features of the magnet
are considered to be merits for the solar axion search.
1) Since only electrical power is needed to cool down,
troublesome cryogen pipes are not necessary.
2) The magnet is equipped with a persistent current switch,
which enables us to remove the power cables
after the excitation of the magnet.
The switch also enables a long term operation.
3) It is possible to incline the magnet
between $-$30 degree and +30 degree.

The X-ray detectors are nine PIN photodiodes.
The size of one diode is $11\times 11 \times 0.5$\,mm$^3$.
The chips for S3590-06 of Hamamatsu photonics
were mounted on special teflon substrate with low radio activity.
Each diode has a charge-sensitive preamplifiers.
They are mounted on a oxygen-free copper plate,
which is cooled down to around 50\,K
by the first stage of the refrigerators.
Between the diodes and plate, acrylic sheets were inserted
to shield X rays from the copper plate.
They are also useful to control the temperature of the FET's
that were used in the head part of the preamplifiers.

The background of the detector is of great importance on sensitivity.
It is about three orders of magnitude
lower than that of the older experiment \cite{Lazarus}.
To reduce the background,
we used lead with 15\,cm thickness,
oxygen-free copper plates with 1\,cm thickness,
and acrylic sheets with 5\,mm thickness.
The lead is kept at room temperature
but the copper plate and acrylic sheets were kept
around 50\,K. The copper plates were also used to cool down
the PIN photodiode array and the head part of the preamplifiers.

The turntable enables long exposure time.
It is noteworthy that a part of the superiority to the older
experiment \cite{Lazarus} comes from this turntable.
The turntable is equipped with two motors to move the magnet
along elevational and azimuthal direction.
The motors are controlled via a CAMAC bus.
Two rotary encoders are equipped for the two axes
to control the direction of the magnet.
The output of the encoders are read via the CAMAC bus.
The magnet can be moved almost 360$^\circ$
along the horizontal direction
and $\pm 28^\circ$ along the elevational direction.
The horizontal and elevational origin
was determined by a precision theodolite.
We determined the north direction by observing a star,
$\beta$-Ori, which is usually called Rigel.
The horizontal level was determined by a split level.

Since we obtained null result,
it is important to estimate systematic errors
to direct the detector toward the sun.
We considered many origins that might cause
the systematic errors and concluded
the errors are smaller than $\pm$0.50\,mrad in the azimuthal direction
and $\pm$0.45\,mrad in the elevational direction.
The principal contribution to the error for the azimuthal direction
is due to the inaccuracy to determine
the north direction, $\pm$ 0.47\,mrad.
That for the elevational direction is due to the table
fluctuation when it rotates, $\pm$0.30\,mrad.
In calculating the position of the sun,
we used the U. S. Naval Observatory Vector Astronomy
Subroutines \cite{NOVAS}.
The location of the detector is
at 139$^\circ$46'1''\,E
and 35$^\circ$42'37''\,N.

The output signals of the preamplifiers are stored
and analyzed to obtain energy spectrum of the signals.
They are fed into Flash Analog-to-Digital Converters
(FADC's, REPIC RPC-081) and shaping amplifiers (Clear Pulse CP4026).
FADC's recorded waveforms of the output signals
over 50\,$\mu$sec before and after a trigger
with a sampling period of 0.1\,$\mu$sec.
The trigger is generated by discriminating
the signals of the shaping amplifiers.

Since we used an array of PIN photodiodes
and the trigger generator, we must evaluate
absolute detection efficiency for converted X rays,
an effective area, and trigger efficiency.
These factors were evaluated by examining a PIN photodiode,
the array, and the total detector system
including the trigger generator.

First, we examined the absolute detection
efficiency of a PIN photodiode.
It is not easy to measure the efficiency over the 2--20\,keV region,
where the axion signal is expected to be most prominent.
Therefore we estimated a lower bound
of the efficiency for the energy region.
We irradiated the central part of a diode by X rays from $^{55}$Fe,
whose intensity was calibrated.
We found the absolute efficiency
for 5.9\,keV X rays is $1.16\pm 0.15$.
We assumed that the inefficiency of the measurement,
if any, is due to a dead surface layer of the PIN photodiode.
We found at 95\% confidence level that
the efficiency for 5.9\,keV X rays is larger than 0.8 and
the dead surface layer is thinner than $6.1\,\mu$m.
We assumed the thickness of the dead surface layer is $6.1\,\mu$m
and calculated the absolute efficiency over the 2--20\,keV region.
For example, $E_{PIN}(4\,{\rm keV}) = 0.5$,
$E_{PIN}(6\,{\rm keV}) = 0.8$, and $E_{PIN}(10\,{\rm keV}) = 0.9$.
We define this theoretical curve as $E_{PIN}(e)$.

We also scanned the area of a PIN photodiode
by a collimated $^{55}$Fe X-ray source to measure the effective area.
As a result, we found $9\times 9\,\rm mm^2$ region
was a conservative estimate of the effective area.

Next, we examined the array of the nine diodes.
We took a photograph of the array
and determined their exposed area.
The corresponding inefficiency is determined to be 0.03.
In addition, there are inefficiency
associated with the inaccuracy of the direction
of the detector toward the sun.
This inefficiency was estimated with the values
given before and found to be less than 0.14.
Thus the overall inefficiency associated
with these systematic effects becomes $D=0.17$

Finally, we confirmed the trigger efficiency
in the energy region 4--20\,keV is 100\%.
We must measure trigger efficiency
because the outputs of the shaping amplifier are discriminated.
We measured them by varying amplitudes
of test pulses to preamplifiers
stepwise ranging from zero to the voltage
which corresponds to the full scale energy.
Since all data in the energy region 4--20\,keV passed the event selection,
which is described in the following section,
we concluded the trigger efficiency in the energy region
4--20\,keV was 100\%.

\section{Measurements and Analysis}
We have performed a search for solar axions
from 26th till 31st December 1997.
Here, we describe a summary of the data and
a method of the analysis.

We classified the livetime into three
categories by position of the sun.
1)~While the detector axis is tracking the sun,
the livetime is counted as source run time, 
$1.90\times 10^5$\,sec.
2)~When the axis is not tracking the sun,
it is added to background run, 
$1.98\times 10^5$\,sec.
3)~For the rest of the time,
it is added to the grey time, 
$2.4\times 10^3$\,sec.

Event selection cuts are designed to discard cosmic muon events;
saturated events and cross talk events are cut.
The cross talk is not a problem
because the cross talk to other channels are so weak
that low energy events from axions are
not systematically discarded by this cut.
The inefficiency from random coincidence is measured along with
the trigger efficiency and found to be negligible.

We need to shape the recorded waveform by a software program,
since we recorded the waveform of the preamplifier output.
We used the cusp shaping
because undershooting is not a problem in our situation.
We selected a time constant to be 6\,$\mu$sec
to get the best $S/N$ ratio.

Energy calibration was done by
irradiating X or $\gamma$ rays from several radioactive sources
before the detector installation.
After the installation, we can use test pulses to
checke stability of the electronics of the detectors
and the trigger efficiency over the measurement.

\begin{figure}
  \begin{center}
    \leavevmode
    \epsfxsize=15cm
    \epsffile{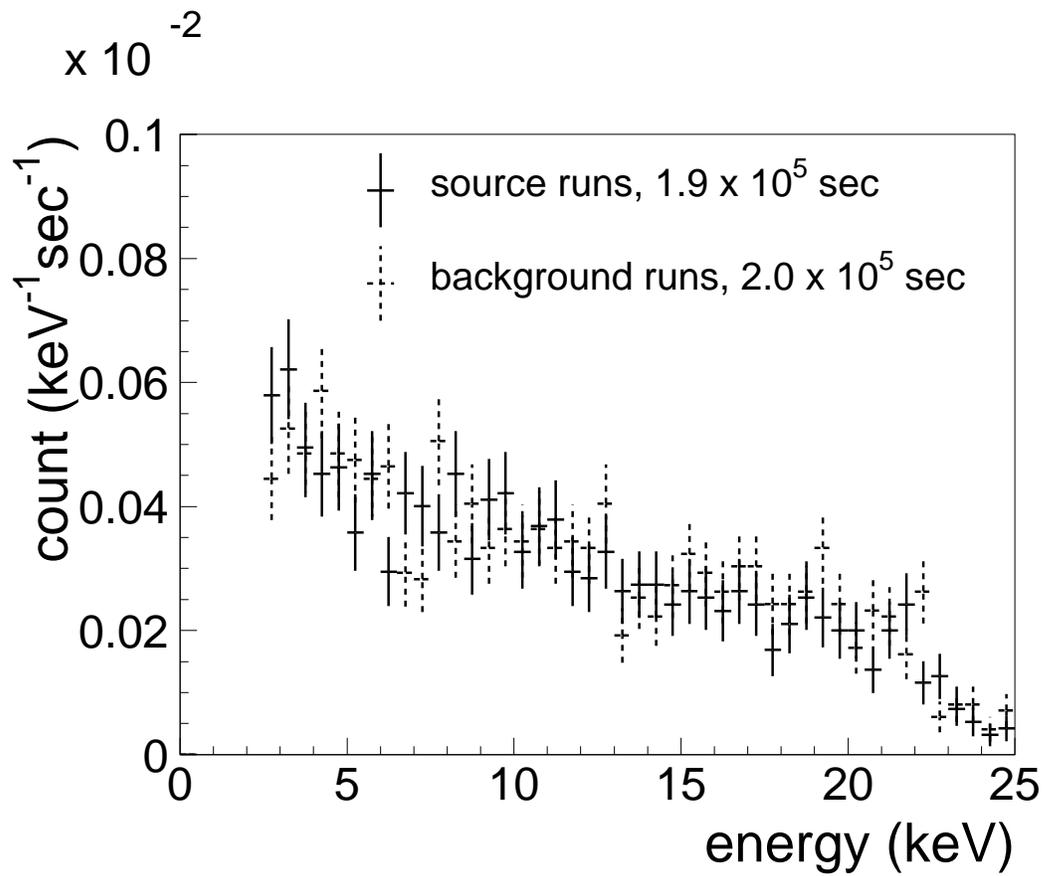}
    \caption{Energy spectrum of the source runs and the background runs.
      Each spectrum is obtained by summing up the data of
      nine PIN diodes.}
    \label{fig:spectrum}
  \end{center}
\end{figure}

From the data, we finally obtained energy spectra
of source runs and background runs.
Fig.\ \ref{fig:spectrum} shows $S_s(E)/t_s$ and $S_b(E)/t_b$,
where $S_s(E)$ is the energy spectrum
of source runs, $S_b(E)$
is the energy spectrum of background runs,
$t_s$ is the livetime of source runs, and
$t_b$ is the livetime of background runs.

Since the signal from the axion can be observed
only in the source run, we subtracted background
contribution from it and fit the theoretical axion signal to it.
The signal strength is proportional to $g_{a\gamma\gamma}^4$.
We observed that $g_{a\gamma\gamma}$ is consistent with zero
and we gave a limit to it.

First, conversion probability from axions to photons is calculated.
In Ref.\ \cite{Bibber89}, the probability is calculated.
In our experimental situation,
effective photon mass $m_\gamma=0$ and
attenuation coefficient $\Gamma=0$
because the conversion region is vacuum.
The conversion probability, $p_{\gamma}$,
is written by a Fourier transformation of the magnetic field:
\begin{eqnarray}
  p_\gamma(E, x, y) &=& \left|\int^L_0\frac{B(x, y, z)}{2M}
  \exp\left(izq\right)dz \right|^2,
\end{eqnarray}
where $B$ and $L$ are the magnetic field and its length,
$q=m_a^2/2E$ is the momentum transfer,
$z$ is a coordinate along the magnet axis,
and $x, y$ are coordinates perpendicular to the $z$ axis.
This provides a spectrum that is expected in our experiment: 
\begin{eqnarray}
  S_{expected}(E) &=&  \sum_i \int_{A_i} dxdy \int_0^\infty de
  \frac{d\Phi_a(e)}{de}
  p_\gamma(e, x, y) \nonumber\\
  {} && \times E_{PIN}(e)
  \exp(-(e-E)^2/2\sigma_d^2)E_{trig}(E)(1-D)\nonumber\\
  {} &\simeq& \sum_i A \int_0^\infty de
  \frac{d\Phi_a(e)}{de}
  p_\gamma(e, x_i, y_i) \nonumber\\
  {} && \times E_{PIN}(e)
  \exp(-(e-E)^2/2\sigma_d^2)E_{trig}(E)(1-D),
  \label{eq:totspec}
\end{eqnarray}
where $d\Phi_a(e)/de$ is the differential flux of the axion \cite{Avignone},
$x_i$ and $y_i$ are the coordinates of the center of
the $i$-th PIN photodiode
and $A$ is the area of the PIN photodiode.
$E_{PIN}(e)$ is the efficiency of the PIN photodiodes.
$E_{trig}(E)$ is the trigger efficiency.
In Eq.\ (\ref{eq:totspec}), we assumed that the trigger
efficiency is a function of a shaped pulse height.
$\sigma_d$ is the energy resolution of the X-ray detectors.
The energy resolution for each detector was determined by test pulses.
Since we use summed spectrum over the nine PIN diodes,
an error of energy calibrations must be added to the resolution.
We estimated the effects and concluded
that $\sigma_d = 0.34$\,keV.
$D$ is the inefficiency defined
in the section \ref{sec:experimentalapparatus}.

If we fix $m_a$, a theoretical spectrum of converted photons,
$S_{expected}(E)$, can be calculated.
The shape of the spectrum differs by $m_a$
because the Fourier component of the magnetic field
distribution is not constant.
We varied the height of the theoretical
spectrum for a value of fixed $m_a$
and fit it to the difference of the spectra
shown in Fig.\ \ref{fig:spectrum}.
The fitting region is restricted to the energy region
4--14\,keV because the trigger efficiency is almost 100\%
and most events are concentrated in the region.

The fit was calculated using the MINUIT of the CERN Program Library
\cite{MINUIT}. Fitting was repeated with various values for $m_a$
in the range 0.001--1\,eV.
All heights are consistent with zero within statistical errors and
thus we constrain $g_{a\gamma\gamma}$ for each value of $m_a$.

\begin{figure}
  \begin{center}
    \leavevmode
    \epsfxsize=15cm
    \epsffile{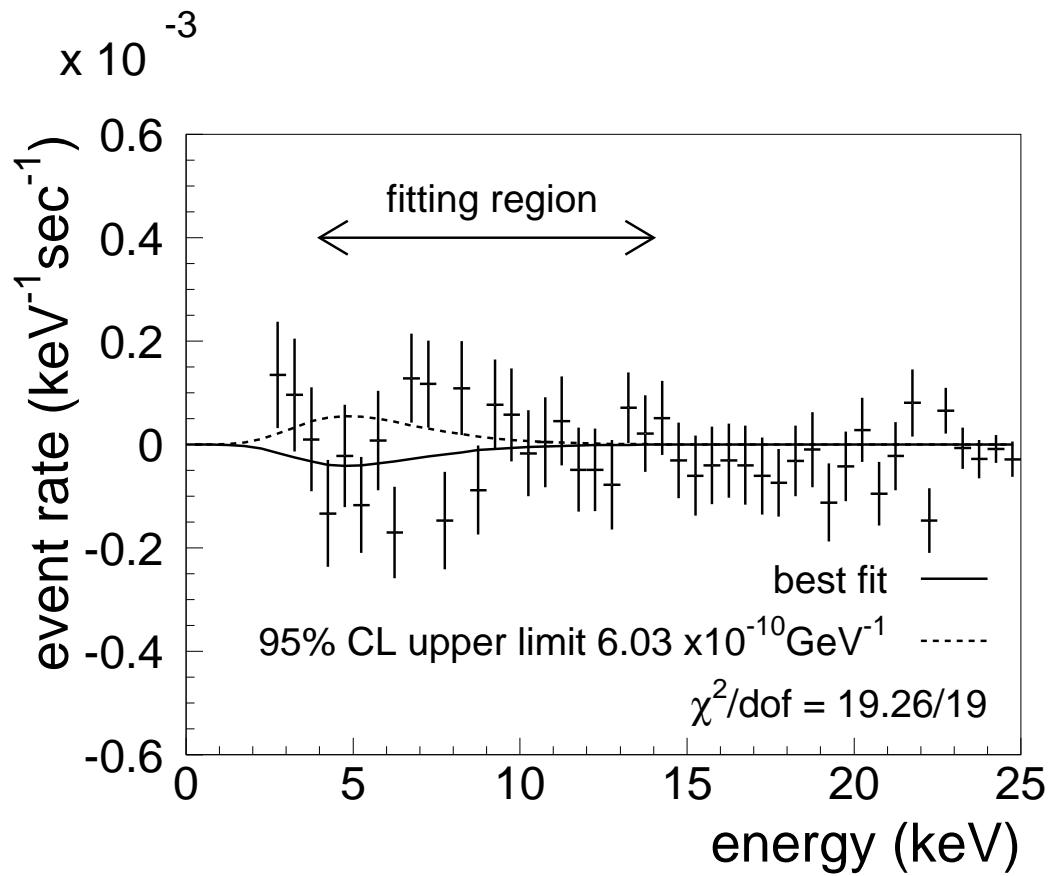}
    \caption{Fitting of the axion spectrum with experimental data.
      $m_a=0.001$\,eV is assumed.}
    \label{fig:fit}
  \end{center}
\end{figure}

\begin{figure}
  \begin{center}
    \leavevmode
    \epsfxsize=15cm
    \epsffile{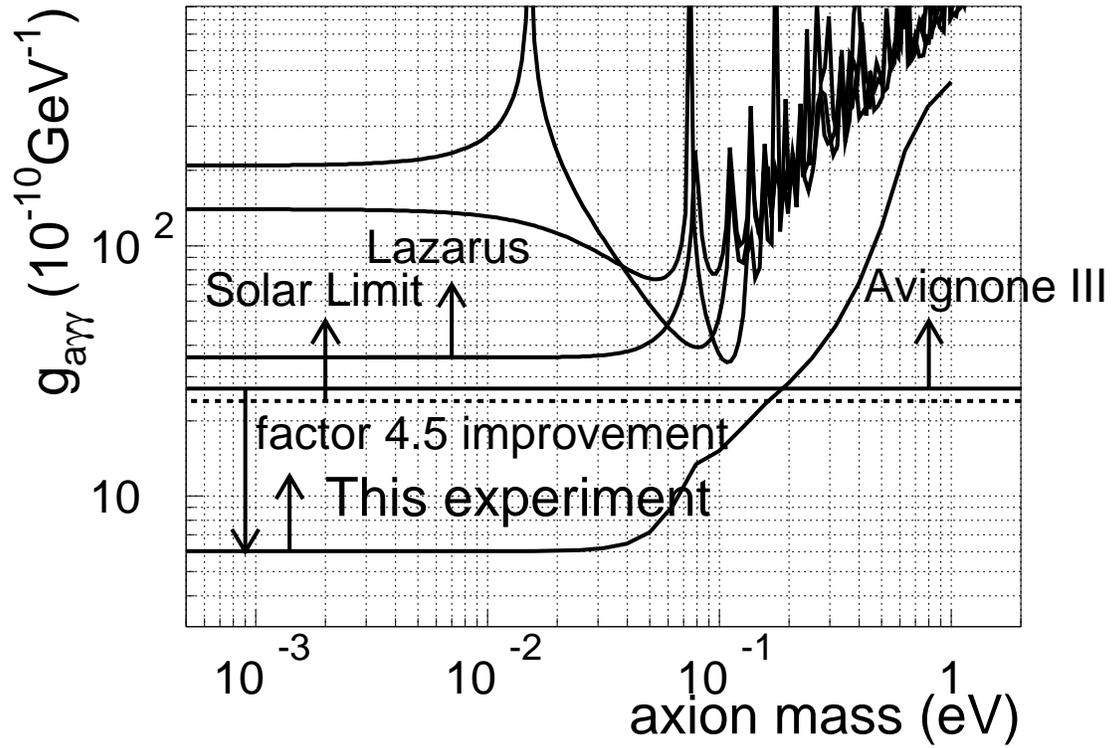}
    \caption{The 95\% CL upper limit for the two-photon coupling
      constant $g_{a\gamma\gamma}$. The result of Lazarus
      \protect\cite{Lazarus} is our calculation.
      He corrected his result \protect\cite{privatecom}.
      The recent result of Avignone III was
      obtained with a germanium detector \protect\cite{Avignone}.}
    \label{fig:limit}
  \end{center}
\end{figure}

For example, the result of the fit with $m_a=0.001$\,eV
is shown in Fig.\ \ref{fig:fit}.
The solid lines correspond to the best fit,
and the dashed lines correspond to the 95\%
confidence level upper limit. 
We obeyed the method of Particle Data Group
\cite{PDG} to obtain the limits.
Since the height of the spectrum is positive definite,
we integrate the probability on the positive side
and find the relative 95\% CL upper limit.
Thus we obtained 95\% CL upper limits for various
$m_a$ in the range 0.001--1\,eV
as plotted in Fig.\ \ref{fig:limit}.

We checked stability of the event rates.
The event rate was constant within statistics
through out the experiment. No dependence on the direction
of the detector is observed either.
The event rate in the energy region 4--14\,keV for each detector
were examined and found no significant difference among detectors.

\section{Conclusion}
We have searched for axions which could be produced
in the sun by exploiting their conversion to X rays
in a static magnetic field.
The signature of the solar axion
would be an increase in the rate of X rays
detected in the magnetic helioscope
when the sun is within its acceptance.
We performed solar axion search for five days
with nine PIN photodiodes as X-ray detectors.
Since the signal from the axion could be observed
only during the source runs, we subtracted background
contribution from the data and fit the axion signal spectrum to them.
Fitting was repeated with various values for
$m_a$ in the range 0.001--1\,eV. As a result,
we found the axion signal is
consistent with zero within statistical errors
for any $m_a$ in this range.
We set 95\% confidence limits on the axion coupling
to two photons as $g_{a\gamma\gamma} \equiv 1/M < 6.0\times
10^{-10}$\,GeV$^{-1}$, provided the axion mass $m_a < 0.03$\,eV.
This is the first experiment which has sufficient sensitivity
to explore the axions whose coupling constant $g_{a\gamma\gamma}$
is smaller than the solar limit, $2.4\times 10^{-9}$\,GeV$^{-1}$.
The limit on the coupling is factor 4.5 more stringent
than recent experimental result in the region $m_a < 0.03$\,eV.

\section*{Acknowledgements}
We are indebted to S.~Mizumaki who fabricated
the persistent current switch of the magnet.
We are also thankful to Y.~Sato for her support
in operating the superconducting magnet
in the beginning stage of the experiment.
We further thank S.~Otsuka for
his support of our engineering work.
We express gratitude to F.~Shimokoshi
who designed the original form of the preamplifier circuit.
We express gratitude to Y.~Makida and K.~Tanaka
who helped us to prepare various instruments associated with the magnet.
We are indebted to H.~Hara for his contribution
in the first stage of the development of the X-ray detectors.

This research is supported by the Grant-in-Aid for COE Research
by the Japanese Ministry of Education, Science, Sports and Culture,
and also by the Matsuo Foundation.

\end{document}